\documentclass[preprint2,fleqn]{aastex}

\usepackage{graphicx}
\usepackage{amsmath}
\slugcomment{}

\shorttitle{X-ray emission from HH~248}
\shortauthors{L\'opez-Santiago et al.}

\begin{document}


\title{X-ray emission from stellar jets by collision against high-density molecular clouds: an application to HH~248}


\author{J. L\'opez-Santiago\altaffilmark{1}, 
R. Bonito\altaffilmark{2,3}, 
M. Orellana\altaffilmark{4}, 
M. Miceli\altaffilmark{3}, 
S. Orlando\altaffilmark{2}, 
S. Ustamujic\altaffilmark{1},
J. F. Albacete-Colombo\altaffilmark{5}, 
E. de Castro\altaffilmark{6}, 
A. I. G\'omez de Castro\altaffilmark{1}}


\altaffiltext{1}{S. D. Astronom\'ia y Geodesia, Facultad de Ciencias Matem\'aticas, 
Universidad Complutense de Madrid, 28040 Madrid, Spain}
\altaffiltext{2}{Dipartimento di Fisica e Chimica, Universit\`a di Palermo, Piazza del Parlamento 1, 
90134 Palermo, Italy}
\altaffiltext{3}{INAF-Osservatorio Astronomico di Palermo, Piazza del Parlamento 1, 
90134 Palermo, Italy}
\altaffiltext{4}{Sede Andina de la Universidad Nacional de R\'io Negro, Argentina}
\altaffiltext{5}{Sede Atl\'antica de la Universidad Nacional de R\'io Negro, Don Bosco y Leloir s/n, 8500 Viedma RN,ÊArgentina.}
\altaffiltext{6}{Dpto. de Astrof\'isica y CC. de la Atm\'osfera, Facultad de F\'isica, 
Universidad Complutense de Madrid, 28040 Madrid, Spain}


\begin{abstract}
{We investigate the plausibility of detecting X-ray 
emission from a stellar jet that impacts against a dense molecular
cloud. This scenario may be usual for classical T Tauri stars with
jets in dense star-forming complexes. We first model the impact of
a jet against a dense cloud by 2D axisymmetric hydrodynamic
simulations, exploring different configurations of the ambient 
environment. Then, we compare our results with XMM-Newton observations
of the Herbig-Haro object HH 248, where extended X-ray emission
aligned with the optical knots is detected at the edge of the nearby
IC 434 cloud. Our simulations show that a jet can produce plasma with
temperatures up to $10^7$ K, consistent with production of X-ray
emission, after impacting a dense cloud. We find that jets denser
than the ambient medium but less dense than the cloud produce
detectable X-ray emission only at the impact onto the cloud. From
the exploration of the model parameter space, we constrain the
physical conditions (jet density and velocity, cloud density) that
reproduce well the intrinsic luminosity and emission measure of the
X-ray source possibly associated with HH 248. Thus, we suggest that
the extended X-ray source close to HH 248 corresponds to the jet
impacting on a dense cloud.}
\end{abstract}





\section{Introduction}
\label{intro}


Classical T Tauri stars (CTTSs) are characterized by being surrounded by 
a disk of gas as a result of the conservation of angular momentum during the star-formation process. 
The central star accretes material from the disk through magnetic funnels \citep{koe91}.
In addition, dense gas from the inner region of the disk is collimated into a jet as explained in 
the context of the widely accepted theory of magneto-centrifugal launching \citep{fer06}.
{Recently, magneto-hydrodynamic (MHD) simulations have been performed to evaluate the 
importance of the accretion disk magnetic field in the launching process of a stellar jet
\citep{mat08,mat09,stu14,sta14}.}
The role of the magnetic field for the collimation of stellar jets in laboratory plasma experiments has been
recently explored by \citet{alb14} through detailed numerical simulations and comparison with observations.

When escaping from the stellar system, the jet moves through circumstellar and interstellar material, 
{interacting with the ambient medium and producing shocks}.
%
{Usually, stellar jets are revealed in the optical by the presence of a chain of knots, 
the so-called Herbig-Haro (HH) objects.} These knots are known to be {associated with} the jet's shock front 
and post-shock regions. {Supersonic shock fronts and post-shock regions along the jet are detected in a wide 
wavelength range, from radio to optical band \citep[][and references therein]{rei01} and may 
be detected also in the ultraviolet {\citep[e.g.][]{gom11,cof12,sch13}.} 
The knotty structure of HH objects within the jet axis is interpreted as the consequence of the pulsing 
nature of the ejection of material by the star \citep[see][]{bon10a}.}
{A possible explanation for this pulsing nature is the variable nature of the stellar wind
\citep{mat09}, whether through stellar magnetic cycles or by variations in the accretion rates}.

{\citet{pra01} {observed} high energy emission from high speed HH jets. In particular, hydrodynamic (HD) 
models} predict X-ray emission from mechanical heating due to shocks produced by the interaction between 
the jet and the ambient medium \citep{bon04}. This is particularly true when the jet is less dense
than the ambient medium \citep{bon07}. Instead, X-ray emission from heavy jets, i.e. jets that 
are denser than the ambient, is {weaker.} 
The analysis of X-ray data from stellar jets can yield constraints to initial conditions of the process 
such as the jet velocity and density. 
However, the low statistics in the few stellar jets detected in X-rays so far makes it difficult to 
achieve robust results \citep[see][for an example]{gud05}.  

X-ray emission from protostellar jets was firstly reported by \citet{pra01}, who detected
an X-ray source {coincident with HH~2} that was not {associated with} any 
other galactic or extragalactic source \citep[see][]{vel14}. Later, \citet{fav02} detected X-ray 
emission from HH~154, {a jet associated with} the protostar L1551~IRS5 in the 
Taurus Molecular Cloud \citep[see also][]{bal03,bon04,bon11,fav06}. Since then, the 
number of X-ray detections from protostellar jets {has been increased:
HH~80/81 \citep{pra04}, TKH~8 \citep{tsu04}, HH~540 \citep{kas05},
HH~210 \citep{gro06}, HH~216 \citep{lin07}, HH~168 \citep{pra09}.}

Several jets from other young stellar objects (YSO) were detected in X-rays, too. 
\citet{gud05} reported likely detection from a jet from the CTTS DG~Tau 
{\citep[see][for recent results]{gud08,sch13}.}
{\citet{ski11} detected evidence for extended X-ray structure in RY Tau.
Finally, X-ray emission from a jet arising from the FU Ori type star Z~CMa was detected by 
\citet{ste09}.} A list of X-ray detections and properties of stellar jets is given in \citet{bon10b}.
The analysis of the X-ray spectrum of stellar jets divides them 
into two classes \citep{bon10b}: (1) X-ray sources detected at the base of the jet (within 2000~AU 
from the stellar source), with temperature $> 2 \times 10^6$~K and (2) X-ray sources detected at 
large distances from the jet source (several thousands of AU), characterized by high luminosity 
and low temperature. The interpretation of this behavior is still not clear, but the difference between 
both classes seems to be related to the location of the X-ray emission.
%
 
{In this work, we study a different scenario for X-ray emission observed from a jet: a stellar jet that moves 
through the interstellar medium (ISM) and impacts a molecular cloud (wall) with a density several times higher 
than that of the ambient. In this scenario, the jet is heavier than the ISM but lighter than the wall. We perform 
simulations with different initial conditions, widely exploring the parameter space. To test our results, we 
compared them with the jet HH~248, detected in X-rays with the XMM-Newton mission 
\citep[Src.~12 in][]{lop13}.}
%

{The organization of this article is as follows. In Section~\ref{model}, we present the numerical 
setup for our simulations. Results for the different initial conditions explored by us are discussed 
in Section~\ref{simulations}. We then study the case of HH~248. A brief description of the system 
is carried out in Sections~\ref{struct}. Specific simulations for HH~248 
are performed in Section~\ref{hh248}. Final discussion and conclusions are presented in 
Section~\ref{discussion}.}

\section{Hydrodynamic model and numerical setup}
\label{model}

\begin{figure}[!t]
\centering
{\includegraphics[width=5cm]{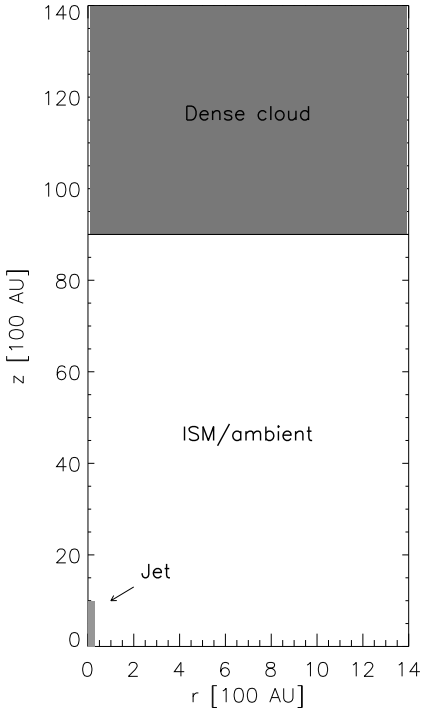}}
\caption{Schematic of the simulation domain setup. Boundary condition 
at the left border is reflective/axisymmetric. Outflow conditions are permitted in the remaining 
boundaries. 
\label{cartoon}}
\end{figure}

{Our model describes the impact of a protostellar jet against a dense
molecular cloud. We adopted the hydrodynamic model discussed by
\citet[][hereafter BOP07]{bon07}, extended to include an
ambient environment characterized by the presence of a dense cloud.
The jet impact is modeled by numerically solving the time-dependent
hydrodynamic equations of mass, momentum, and energy conservation
in a 2D cylindrical coordinate system $(r,z)$, assuming axisymmetry
(see BOP07 for details). The model includes the radiative losses
from optically thin plasma, and the thermal conduction, including
the effects of heat flux saturation.  The model is implemented using
the FLASH code \citep{fry00}, with the Piecewise Parabolic
Method (PPM) solver that is better adapted for compressible flows
with shocks \citep{col84}.}

{The jet is assumed to propagate along the symmetry axis, 
$z$, through an unperturbed ambient medium with a dense molecular 
cloud. The cloud is described by a higher density region (wall) in pressure
equilibrium with its surrounding and is situated at $\sim 9000$ AU
from the base of the domain (see Figure~\ref{cartoon}). We chose this distance
as we want to compare the results from the simulations with the
observations of HH 248 (see Section~\ref{hh248} for more details).  The
computational domain extends $\sim 1400$ AU in the $r$ direction
and $\sim 14000$ AU in the $z$ direction. At the coarsest resolution,
the adaptive mesh algorithm used in the FLASH code \citep[PARAMESH;][]{mac00}
uniformly covers the 2D computational domain with an initial 
mesh of $1 \times 10$ blocks, each with $8^2$ cells. We
allow for five levels of refinement, with resolution increasing
twice at each refinement level. The refinement criterion adopted
\citep{loh08} follows the changes in density and temperature.
This grid configuration yields an effective resolution of $\sim 7$~AU 
at the finest level, corresponding to an equivalent uniform mesh of 
$128 \times 1280$ grid points.}

\begin{figure}[!t]
{\includegraphics[width=\columnwidth]{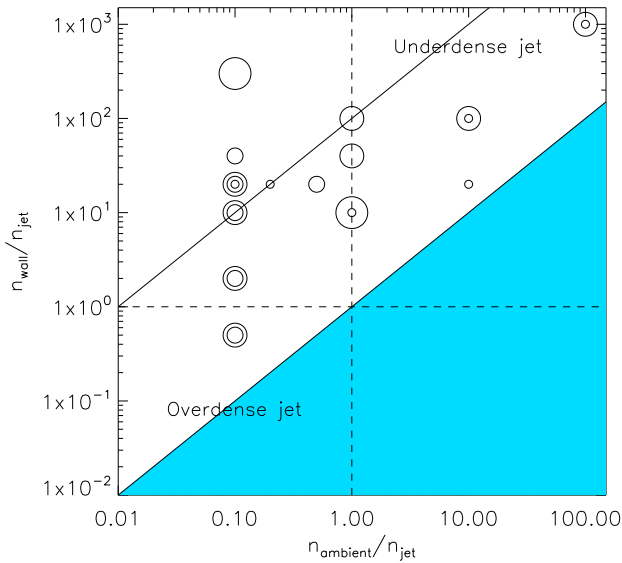}}
\caption{Parameter space of the simulations performed in this work. Small circles correspond
to jet's velocities $v_\mathrm{jet} \leq 500$~km\,s$^{-1}$, medium-size circles are for 
$500 \leq v_\mathrm{jet} < 1200$~km\,s$^{-1}$, and large circles are for simulations with 
$1200 \leq v_\mathrm{jet} < 2000$~km\,s$^{-1}$ and $v_\mathrm{jet} \geq 2000$~km\,s$^{-1}$, 
respectively. The shaded area is the scenario with a molecular 
cloud less dense than the ambient, which is not investigated in this article. The continuous
line corresponds to $n_\mathrm{wall}/n_\mathrm{ambient} = 100$.
\label{spacepar}}
\end{figure}

{We explore the parameter space defined by the density ratio between
the ISM and the jet ($n_\mathrm{ambient}/n_\mathrm{jet}$), the
density ratio between the wall and jet ($n_\mathrm{wall}/n_\mathrm{jet}$),
and the jet's velocity ($v_\mathrm{jet}$). The aim is to investigate which
physical conditions result in the production of detectable X-rays
from the stellar jet after impacting the dense cloud \citep[preliminary results in][]{ore12}. 
{The grid of explored parameters is chosen in order to explore in detail the case 
of a jet over-dense in the ISM that becomes under-dense in the cloud. According to 
previous results \citep[e.g.][]{bon07}, this is the most promising scenario for X-ray 
emission. For completeness, we also explore the remaining two possible scenarios
in less detail: the jet being heavier than both the ambient and the molecular cloud 
and the jet being lighter than both the ambient and the molecular cloud.} The initial 
temperature of the jet is fixed to the value $T_\mathrm{jet} = 10^4$ K in all the simulations.
Figure~\ref{spacepar} shows the parameter space for our simulation. The shaded
area represents the region of the parameter space in which the
ambient would be denser than the molecular cloud. This scenario is
not explored in this work. In BOP07, it is shown
that detectable X-ray emission originates from the interaction of
light jets with the ambient medium. Here we explore the scenario
of a jet heavier than the ambient but lighter than the molecular
cloud. This scenario corresponds to the top-left quadrant in the
figure. 
Table~\ref{tabsim} summarizes the parameters of our simulations. The size of
the circles in Figure~\ref{spacepar} corresponds to four different
velocity ranges for the jet: $v_\mathrm{jet} < 500$~km\,s$^{-1}$, 
$500 \leq v_\mathrm{jet} < 1200$~km\,s$^{-1}$,
$1200 \leq v_\mathrm{jet} < 2000$~km\,s$^{-1}$ 
and $v_\mathrm{jet} \geq 2000$~km\,s$^{-1}$,
respectively from smaller to larger circles. In most cases, we
simulated jets with different initial velocities for the same values
of $n_\mathrm{ambient}/n_\mathrm{jet}$ and $n_\mathrm{wall}/n_\mathrm{jet}$.}

\begin{table}[!t]
\begin{center}
\scriptsize
\caption{Summary of the simulations performed in this work. 
\label{tabsim}}
\begin{tabular}{ccccc}
\tableline\tableline
\noalign{\smallskip}
\multicolumn{1}{c}{$n_\mathrm{ambient}/n_\mathrm{jet}$} &  
\multicolumn{1}{c}{$n_\mathrm{wall}/n_\mathrm{jet}$} &
\multicolumn{1}{c}{$Mach$} &
\multicolumn{1}{c}{$v_\mathrm{jet}$\tablenotemark{\dag}} &
\multicolumn{1}{c}{$\log T_\mathrm{max}$\tablenotemark{\ddag}} \\ 
\noalign{\smallskip}
 &  &  & \multicolumn{1}{c}{(km\,s$^{-1}$)}  & \multicolumn{1}{c}{(K)} \\ 
\noalign{\smallskip}
\tableline
\noalign{\smallskip}
    0.1 &     0.5 &  20 &   936 &  6.70 \\
      "   &     0.5 &  30 & 1404 &  6.84 \\
      "   &     2.0 &  20 &   936 &  6.62 \\
      "   &     2.0 &  30 & 1404 &  6.80 \\
      "   &   10.0 &  20 &  936  &  6.57 \\
      "   &   10.0 &  30 & 1404 &  6.84 \\
      "   &   20.0 &  10 &  448  &  5.70 \\
      "   &   20.0 &  20 &  936  &  6.54 \\  
      "   &   20.0 &  30 & 1404 &  6.83 \\
      "   &   40.0 &  20 &  936  &  6.64 \\
      "   & 300.0 &150 & 2000 &  7.33 \\
\noalign{\smallskip}
\tableline
\noalign{\smallskip}
    0.2 &   20.0 &  20 &  662 &  6.04 \\
\noalign{\smallskip}
\tableline
\noalign{\smallskip}
    0.5 &   20.0 &  40 &  800 &  6.41 \\
\noalign{\smallskip}
\tableline
\noalign{\smallskip}
    1.0 &   10.0 &   20 &   296 &  4.70 \\
      "   &   10.0 & 200 & 2960 &  7.18 \\
      "   &   40.0 & 100 & 1480 &  6.95 \\
      "   & 100.0 & 100 & 1480 &  7.08 \\
\noalign{\smallskip}
\tableline
\noalign{\smallskip}
  10.0 &   20.0 &  100 &  448 &  5.10 \\
     "    & 100.0 &  100 &  448 &  5.60 \\
     "    & 100.0 &  300 &1404 &  6.91 \\
\noalign{\smallskip}
\tableline
\noalign{\smallskip}
100.0 &1000.0&   100 &   148 &  4.40 \\ 
     -    &1000.0& 1000 & 1480 &  7.20 \\ 
\noalign{\smallskip}
\tableline
\end{tabular}
\tablenotetext{\dag}{Derived from the Mach number.}
\tablenotetext{\ddag}{Measured inside the dense cloud, at $z \sim 13000$~AU (see\\ Figure~\ref{fig6}).}
\end{center}
\end{table}

\section{Results from simulations}
\label{simulations}


The evolution of the density and temperature of a jet with distinct ambient-to-jet density ratios
($n_\mathrm{ambient}/n_\mathrm{jet}$) when traveling through a homogeneous ambient was investigated 
in detail in BOP07. In such work, it was shown that the case of a light jet ($n_\mathrm{ambient}/n_\mathrm{jet} > 1$) 
with high Mach number ($M_\mathrm{jet}$) is the most favorable for detection of X-rays. A hot ($T > 10^6$~K) and dense 
blob is formed behind the front shock in both light and heavy ($n_\mathrm{ambient}/n_\mathrm{jet} < 1$) jets when 
the jet's velocity is high enough, but in heavy jets this blob is fainter in X-rays and considerably less extended 
(see Figure~8 in BOP07). In fact, the shock velocity required to produce this faint X-ray emission in heavy 
jets is too high compared to observations ($v_\mathrm{sh} > 2000$~km\,s$^{-1}$). At lower velocities, 
the X-ray luminosity of the heavy jet is low (see BOP07). Consequently, detection 
of X-ray emission from heavy jets is not expected. Instead, the jet's luminosity increases rapidly when it 
becomes lighter than the ambient {(see Section~\ref{hh248}, Figure~\ref{fig4})}. 

{To illustrate the previous discussion, we compare the evolution of the mass density and temperature of 
a heavy jet with similar initial conditions but different velocities. Figure~\ref{simulations} shows mass density 
and velocity cuts in the r--z plane for jets with $M = 10, 20, 30$ and $150$, respectively. In the four cases, 
the initial jet temperature is $10^4$~K and the ambient-to-jet density ratio is $n_\mathrm{ambient}/n_\mathrm{jet} = 0.1$. 
The simulation was stopped at $d \sim 8500$~AU in each case, for purposes of comparison.  }


\begin{figure}[!t]
{\includegraphics[width=\columnwidth]{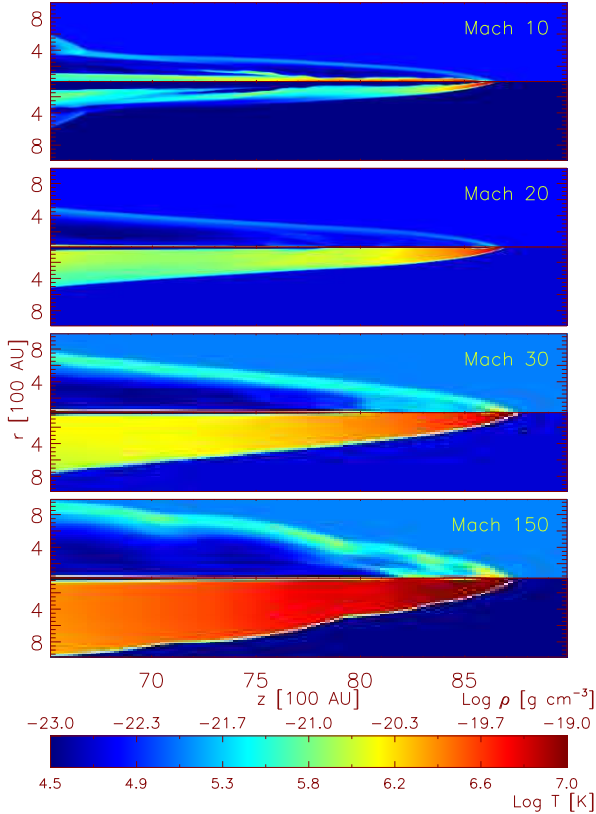}}
\caption{Two-dimensional mass density (upper half-panels) and temperature (lower half-panels) 
cuts in the r--z plane when the jet reaches 8000~AU from the beginning of the simulation. 
All the simulations are for $n_\mathrm{ambient}/n_\mathrm{jet} = 0.1$ but different Mach numbers.
\label{simulations}}
\end{figure}


{Last column in Table~\ref{tabsim} shows the maximum temperature reached by the jet after 
impacting the dense cloud for each simulation. Temperatures similar to those observed
in several X-ray emitting stellar jets (see Section~\ref{intro}) are reached when the velocity 
of the jet is of the order of 700~km\,s$^{-1}$ or higher.
%
}

\begin{figure}[!t]
\centering
{\includegraphics[width=\columnwidth]{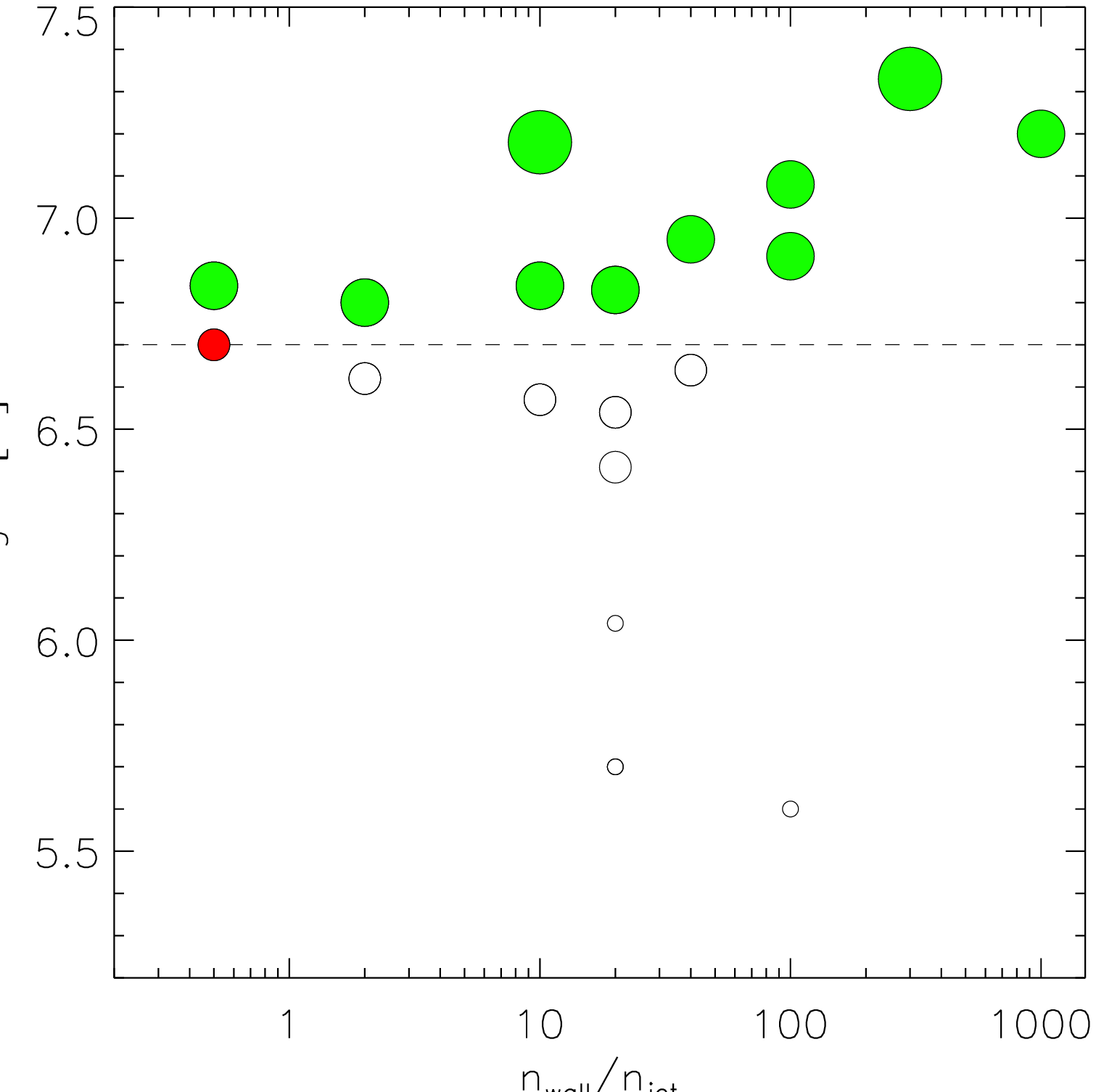}
\includegraphics[width=\columnwidth]{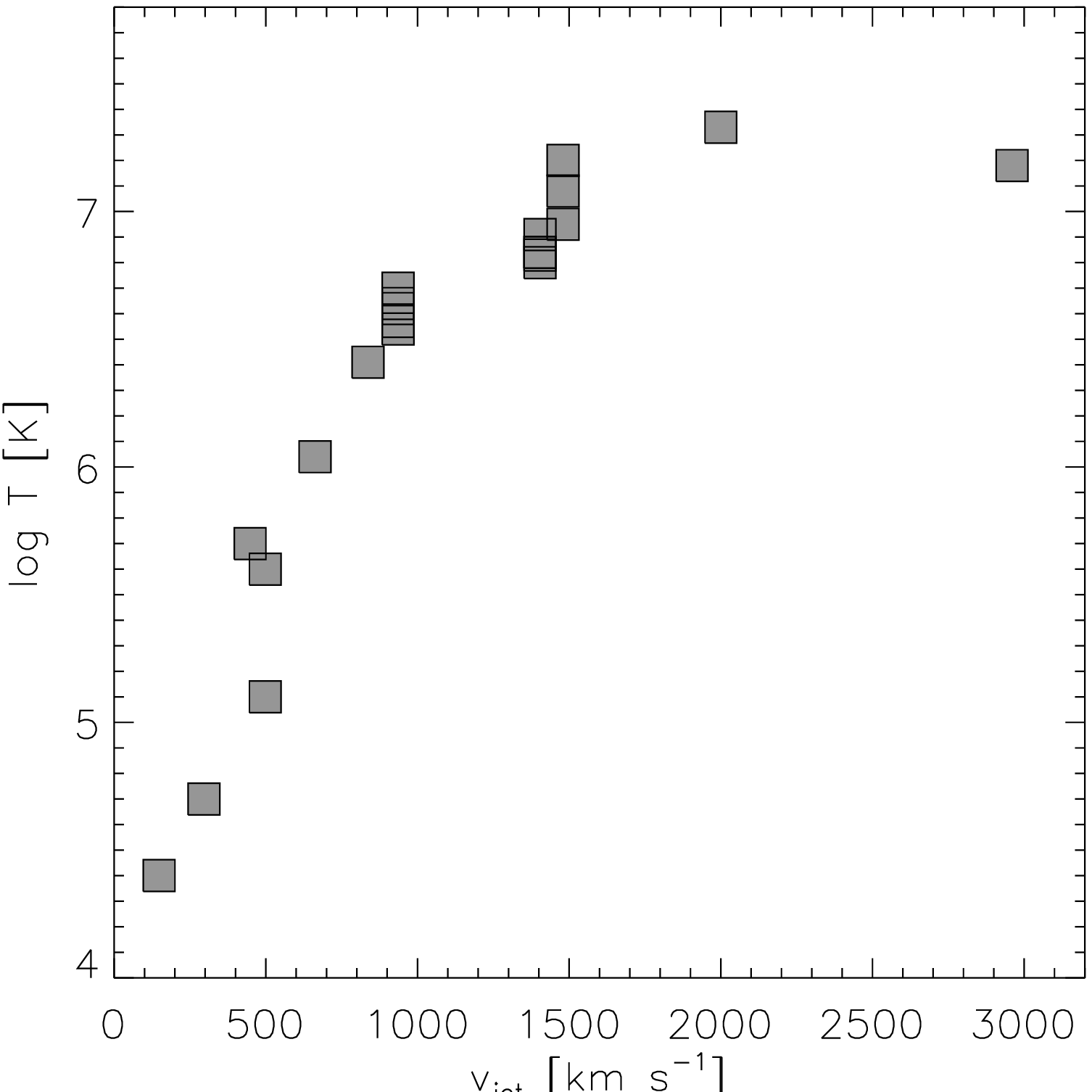}}
\caption{{\bf Top:} Maximum temperature of the jet after impacting the dense cloud for the different simulations, 
compared to the ratio between the jet and the dense cloud densities. Circle sizes are as in 
Figure~\ref{spacepar}. The dashed line indicates a temperature of 5 MK ($kT = 0.43$~keV). 
Filled circles are cases in which the jet reaches temperatures consistent with detected jets 
in X-rays. {\bf Bottom:} Temperature distribution as a function of the velocity of the jet. 
\label{xrayemi}}
\end{figure}

{In Figure~\ref{xrayemi}, we summarize the results of our simulations. The top figure represents
the temperature reached when the jet comes into the dense cloud ($z \sim 120$~AU in 
Figure~\ref{cartoon}) as a function of the ratio between the density of the cloud and the jet.
The bottom panel shows the maximum temperature as a function of the jet's velocity. 
No clear dependence of the temperature with $n_\mathrm{wall}/n_\mathrm{jet}$ is observed (top panel). 
At high density ratios ($n_\mathrm{wall}/n_\mathrm{jet} > 20$), there is only a slight trend of increasing the jet's 
temperature with increasing the density ratio. 
Filled circles in this figure are simulations with $v_\mathrm{jet} \ge 700$~km\,s$^{-1}$. 
%
On contrast, Figure~\ref{xrayemi}, bottom panel shows the temperature of the jet increases rapidly with 
increasing the jet's velocity. Actually, the temperature reaches a plateau at $\sim 1200$~km\,s$^{-1}$ 
where it remains approximately constant ($\sim 10^7$~K) with increasing velocity. For 
$v_\mathrm{jet} \ge 700$~km\,s$^{-1}$, the jet reaches temperatures compatible with X-ray emission 
($T \ge 10^6$~K).}

{Summarizing, a high jet's velocity is responsible for the X-ray emission detected when the 
jet is lighter than the surrounding medium in the proposed scenario.}

\section{HH 248: a candidate for the jet-cloud impact scenario}
\label{struct}

{
The Herbig-Haro object HH~248 is situated to the South-East of the CTTS 
V615~Ori, at the base of the Horsehead Nebula (Barnard~33) and to the South of the NGC~2023 
nebula \citep{mal87}. The detection of an X-ray source at  $\sim 15\arcsec$ 
from HH~248 and $\sim 30\arcsec$ from V615 Ori ($05^\mathrm{h}$ $41^\mathrm{m}$ $25.7^\mathrm{s}$, 
$-02\arcdeg$ $23\arcmin$ $06.0\arcsec$) was reported by \citet{lop13} (their Src. 12). 
The authors did not find any optical or IR counterpart down to 0.1 mJy. Src.~12 of 
\citet{lop13} is classified as an extended source by the \emph{XMM-Newton} SAS source 
detection task, discarding an association with any possible highly absorbed point-like source. 
The \textit{XMM-Newton} observation (ID 0112640201, revolution 419) was centered 
on the reflection nebula NGC~2023. The extended X-ray source was detected $\sim 7\arcmin$ off-axis.
Its proximity to the CTTS V615~Ori may be interpreted as they are related.
However, an analysis of the spectrum of this X-ray source gives $N_\mathrm{H} = 2.3_{-1.4}^{+2.6} 
\times 10^{22}$~cm$^{-3}$ (errors are at 90\% confidence level) for the column density of the source and 
$T \sim 10^7$~K as a lower limit.}
The $N_\mathrm{H}$ value is considerably larger than the one obtained by \citet{lop13} for V615~Ori 
($N_\mathrm{H} = 0.50_{-0.04}^{+0.05} \times 10^{22}$~cm$^{-3}$). We conclude that 
V615~Ori is not the driving source of the extended X-ray emission detected.

\begin{figure}[!t]
\begin{center}
{\includegraphics[width=\columnwidth]{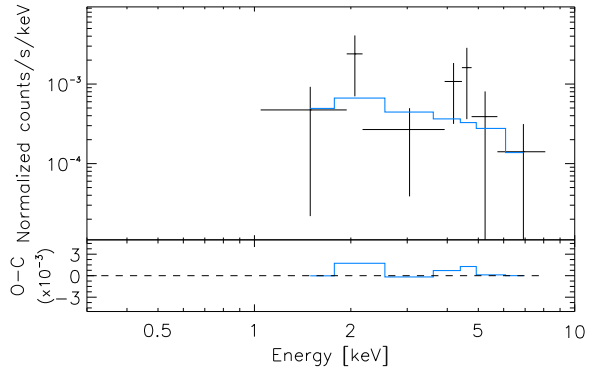}}
\caption{Background-subtracted XMM/EPIC spectrum of the extended source including the 
model fit. 
\label{spec}}
\end{center}
\end{figure}

{Assuming a distance of $\sim 400$~pc for the source, $L_\mathrm{X} = 3 \pm 2 \times 10^{30}$~erg\,s$^{-1}$
and $EM \approx 2 \times 10^{53}$~cm$^{-3}$. This analysis was performed
by ourselves by fitting an absorbed hot plasma model \citep{smi01} to the observed EPIC/PN spectrum 
using the XSPEC spectral fitting package \citep{arn96,arn04} with Cash statistics \citep{cas79}.
{We notice that XSPEC treats both the source and the background spectra separately
when Cash statistics is used.}
The poor statistics of the source does not permit to study the properties of the X-ray emission in 
more detail (see Figure~\ref{spec}).}

\begin{figure}[!t]
\begin{center}
{\includegraphics[width=\columnwidth]{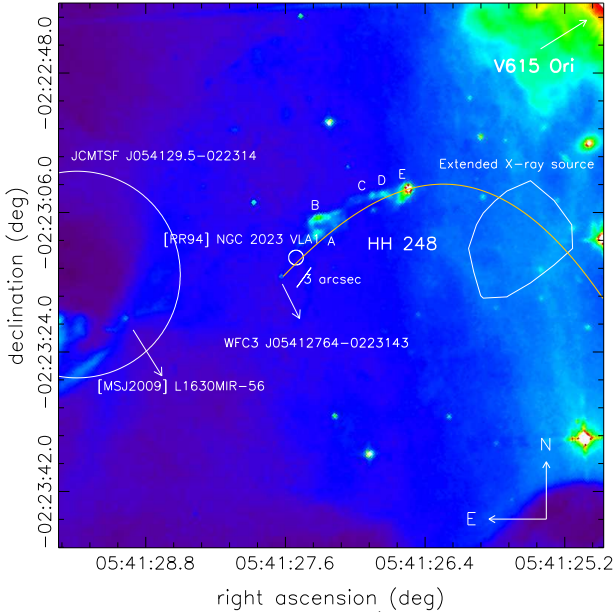}}
\caption{High-contrast HST/WFC3 H band image centered at HH~248 
{(Obs. ID 11548, PI: S. Megeath)}. Different high-density 
knots are identified with capital letters. The circles mark radio sources identified in the literature. 
Their radius corresponds to the 1$\sigma$ positional error. The position of the extended X-ray 
source identified by us as the interaction of jet with the molecular cloud is also marked. 
The continuous line is the precession curve. 
\label{fig1}}
\end{center}
\end{figure}

{In the optical band,} at an angular resolution of ~0.3$\arcsec$, HH~248 shows a knotty structure 
with five clear {blobs}. {Figure~\ref{fig1} shows an HST/WFC3 image in the H band
(\dataset[ADS/Sa.HST\#IB0L9M010]{Obs. ID 11548, obs. date 2010-10-19, exp. time 2496.170~s, 
PI: S. Megeath}).}
A very faint source (WFC3~J05412764-0223143) is detected 3$\arcsec$ to the South-East of the 
radio source [RR94] NGC 2023 VLA1 {reported by \citet{rod94}}, {at the position ($\alpha$, $\delta$) = 
(05$^\mathrm{h}$ 41$^\mathrm{m}$ 27.64$^\mathrm{s}$, -02$\arcdeg$ 23$\arcmin$ 14.3$\arcsec$)}. 
WFC3~J05412764-0223143 is connected to HH~248 by a filamentary 
structure. WFC3~J05412764-0223143 is very likely the driving source of HH~248. 
{Its magnitude in the F160W filter of the HST/WFC3 determined by us (see bellow) is $\sim 23.5$~mag, 
corresponding to a flux $F_\nu = 0.46 \pm 0.01$~$\mu$Jy. Compared to the fluxes determined for the knots
(see Table~\ref{tab1}) this value is between two and three orders of magnitude lower. 
WFC3~J05412764-0223143 must be highly absorbed, with $A_\mathrm{V} > 13$~mag 
according to its observed magnitude in the F160W filter. To determine the fluxes of 
WFC3~J05412764-0223143 and the knots of HH~248, we performed aperture photometry in the 
pipeline processed data were obtained from the Mikulski Archive for Space 
Telescopes (MAST) using the DAOPHOT package \citep{ste87}. An aperture radius of $1\arcsec$ was used 
in each case to prevent contamination from other close knots, whose typical separation is $\sim 2\arcsec$.
We notice that a $1\arcsec$-radius extraction region encircles more than the 95\% of each bright knot even 
for the case of knot E.} Sky background value was fixed for subtraction. This value was 
estimated from the surroundings of the jet. An aperture correction was applied following the 
guidelines of the WFC3 instrument\footnote{http://www.stsci.edu/hst/wfc3/phot\_zp\_lbn.}. 

\begin{table}[!t]
\begin{center}
\caption{WFC3 F160W photometry of the knots and their probable driving source.
\label{tab1}}
\small
\begin{tabular}{lrr}
\tableline\tableline
\noalign{\smallskip}
Source & \multicolumn{1}{c}{AB} & \multicolumn{1}{c}{F$_\nu$} \\
             & \multicolumn{1}{c}{(mag)} & \multicolumn{1}{c}{($\mu$Jy)} \\
\noalign{\smallskip}
\tableline
\noalign{\smallskip}
{J05412764-0223143} & 23.49 $\pm$ 0.18 & 0.46 $\pm$ 0.01 \\
A             & 18.21 $\pm$ 0.01 & 58.96 $\pm$ 0.09 \\
B             & 17.99 $\pm$ 0.01 & 72.17 $\pm$ 0.10 \\
C             & 18.95 $\pm$ 0.02 & 29.88 $\pm$ 0.06 \\
D             & 18.42 $\pm$ 0.01 & 48.71 $\pm$ 0.08 \\
E             & 16.56 $\pm$ 0.01 & 270.16 $\pm$ 0.19 \\
\noalign{\smallskip}
\tableline
\end{tabular}
\end{center}
\end{table}

{The chain of knots in HH~248 follows a curve path typical of precessed jets \citep[e.g.][]{fra14}.
In Figure~\ref{fig1} we show a simple fit of a precession curve \citep[e.g.][]{mas12} to the knots and
the IR source. The curve passes through the X-ray source. Hence, 
in the scenario proposed by us, the X-ray source is the bow shock produced by the encounter with a dense 
cloud of the jet flowing from WFC3~J05412764-0223143.} 

\section{Simulations for HH~248}
\label{hh248}

\begin{table}[!t]
\begin{center}
\scriptsize
\caption{Initial physical conditions in the jet, the ambient (ISM) and the \emph{wall}.
\label{tab3}}
\begin{tabular}{ll}
\tableline\tableline
\noalign{\smallskip}
Parameter &  \multicolumn{1}{c}{Value}  \\ 
\noalign{\smallskip}
\tableline
\noalign{\smallskip}
\multicolumn{2}{c}{Jet} \\
\noalign{\smallskip}
\tableline
\noalign{\smallskip}
Density & 500 cm$^{-3}$ \\ 
Temperature & $10^4$ K \\
Velocity & $1.35 \times 10^3$ km\,s$^{-1}$ \\
\noalign{\smallskip}
\tableline
\noalign{\smallskip}
\multicolumn{2}{c}{ISM} \\
\noalign{\smallskip}
\tableline
\noalign{\smallskip}
Dentity & 50 cm$^{-3}$ \\
Temperature\tablenotemark{a} & $10^5$ K\\
\noalign{\smallskip}
\tableline
\noalign{\smallskip}
\multicolumn{2}{c}{Wall} \\
\noalign{\smallskip}
\tableline
\noalign{\smallskip}
Dentity & $5 \times 10^3$ cm$^{-3}$ \\ 
Temperature\tablenotemark{a} & $10^3$ K\\
\noalign{\smallskip}
\tableline
\end{tabular}
\tablenotetext{a}{Temperature is fixed to maintain pressure equilibrium} 
\end{center}
\end{table}

For our problem, we adopted a density of $5 \times 10^3$~cm$^{-3}$ for the wall as a lower limit, based of 
measurements of the density in the surroundings of the Horsehead Nebula, where HH~248 is located \citep{hab05}. 
{Then we scaled the density of the jet to have a light jet traveling 
through the ISM ($n_\mathrm{jet}/n_\mathrm{ISM} = 10$) and a heavy jet through the wall 
($n_\mathrm{jet}/n_\mathrm{wall} = 0.1$).}
%
Pressure equilibrium is imposed. As a consequence, the temperatures of the three regions are fixed. 
We assume an initial temperature for the jet of $10^4$~K. Thus, the temperatures of the {ISM} and
the wall are, respectively, $10^5$ and $10^3$~K. 
Table~\ref{tab3} summarizes the parameters of our {model}. 
Initially, the jet is defined as a homogeneous cylinder with size $r \sim 100$~AU and $z \sim 1000$~AU. 
{The distance between the jet's base and the dense cloud is kept as in Section~\ref{model}.
The actual projected distance between the X-ray source and WFC3~J05412764-0223143 is 9500-12000~AU, 
assuming a distance from the Earth between 350 and 450~pc as deduced from the measurements for 
$\sigma$~Orionis  \citep[e.g.][]{cab08}.}

\begin{figure}[!t]
{\includegraphics[width=\columnwidth]{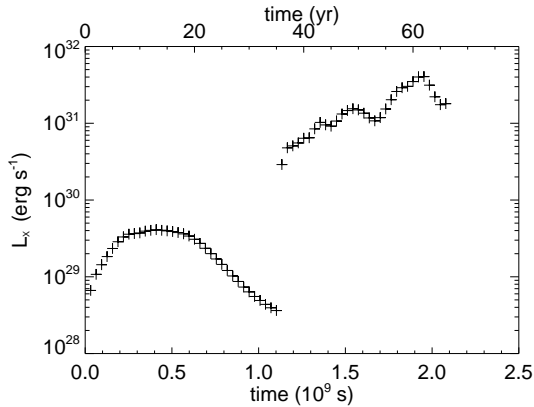}}
\caption{Evolution of the jet's intrinsic luminosity with time in the [0.3-10]~keV energy band. 
The jet impacts onto the denser cloud at $t = 1.1\times 10^9$~s.
\label{fig4}}
\end{figure}

The evolution of the jet's luminosity with time is shown in Figure~\ref{fig4}. X-ray luminosities were 
derived as discussed in \citet{bon10b}. In our space domain, 
the jet reaches the wall after $\sim 35$~years.  Figure~\ref{fig4} shows that the intrinsic luminosity 
of the jet reaches $\sim 10^{31}$~erg\,s$^{-1}$ after its impact against the molecular cloud. However, 
during the first stage of the jet's evolution, while the jet crosses the (less dense) {ISM}, 
its X-ray luminosity remains at a lower value ($\sim10^{29}$~erg\,s$^{-1}$). An X-ray luminosity 
$> 10^{30}$~erg\,s$^{-1}$ for the jet inside the high density molecular cloud is in agreement with 
the unabsorbed X-ray luminosity determined from the fit to the \emph{XMM-Newton} 
data\footnote{The agreement with the observations is satisfactory because we did not perform any 
fine-tuning of the initial conditions which is needed to accurately reproduce the observed luminosity. 
For instance, a better fit might be obtained by decreasing the ambient-to-jet density contrast (see BOP07).} 

\begin{figure*}[!t]
{\includegraphics[width=\textwidth]{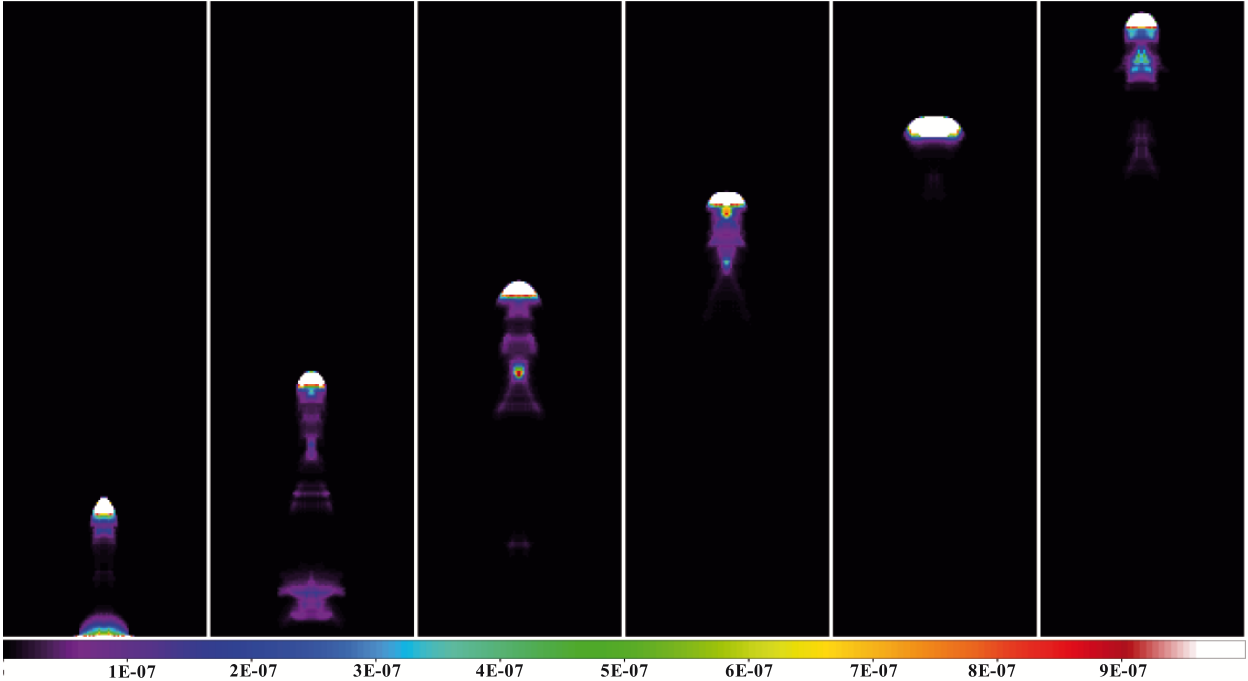}}
\caption{X-ray map of the jet after it reaches the high-density molecular cloud (\emph{wall}). 
The period range is 40--65 yr with a time binning of 5 yr. The pixel size if 1~AU. The color scale 
is $10^{-7}$--$10^{-6}$ cnt\,s$^{-1}$\,pixel$^{-1}$. 
\label{fig5}}
\end{figure*}

\begin{figure}[!t]
\centerline
{\includegraphics[width=4cm]{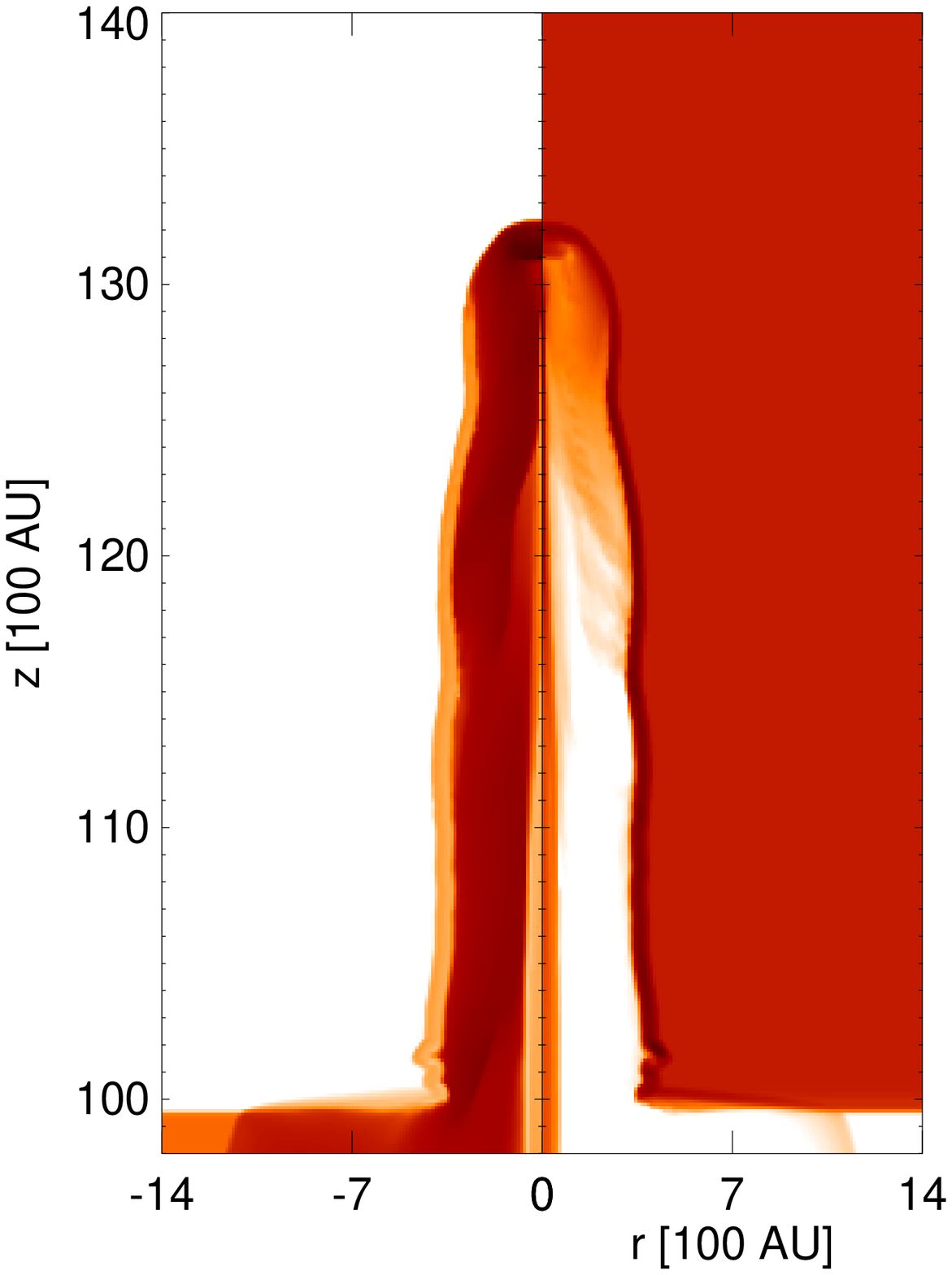}
\hspace{0.35cm}
\includegraphics[width=0.6cm]{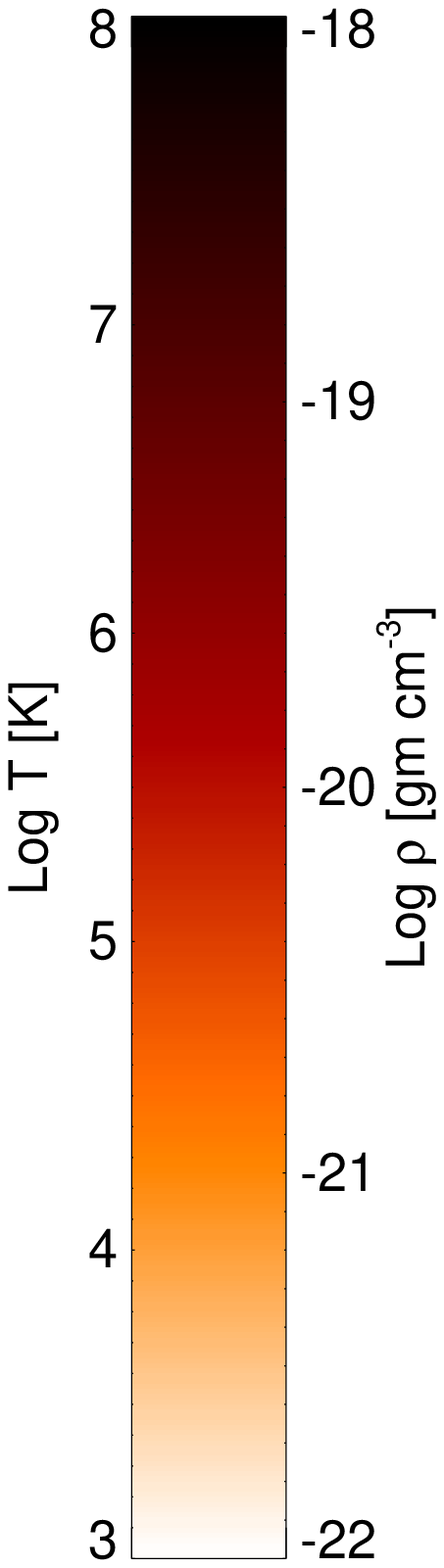}}
\vspace{0.3cm}
\caption{Enlargement of the temperature (left panel) and density (right panel) maps derived 
from our simulations describing the jet/wall interaction (at 45~yr from the beginning of the 
simulation). The complete movie of the evolution is also available as online material.
\label{fig6}}
\end{figure}

Figure~\ref{fig5} shows the X-ray evolution map of the jet as it travels through the 
high-density molecular cloud. Each panel corresponds to a time delay of five years. 
The first panel is for 30 yr from the beginning of the model run. 
{Figure~\ref{fig6} shows the temperature (left panel) and the density (right panel) maps 
describing the jet/wall interaction.}

\begin{figure}[!t]
{\includegraphics[width=\columnwidth]{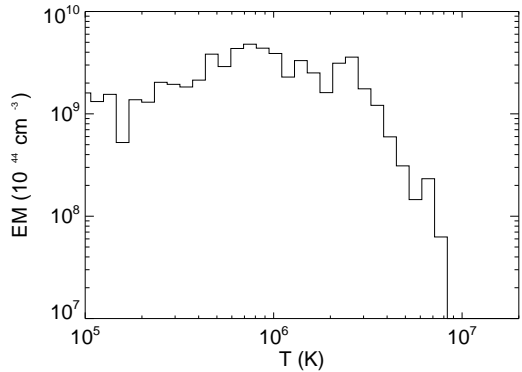}}
\caption{Emission measure distribution ($EM$) as a function of temperature ($T$) for the
penultimate panel of Figure~\ref{fig5}, at a time of 25 yr after the interaction of the jet 
with the molecular cloud (60~yr from the beginning of our simulation). 
\label{fig7}}
\end{figure}

{The emission measure ($EM$) distribution near the end of the simulation (penultimate panel in Figure~\ref{fig5}) 
shows a small bump at $T \sim 10^6$~K (see Figure~\ref{fig7}). The mean value of the $EM$ obtained from our simulations is compatible 
with the $EM$ obtained from the fit to the X-ray data ($\sim 2 \times 10^{53}$~cm$^{-3}$). Note that the $EM(T)$ at 
temperature $< 1$~MK shows a plateau analogous to that found by BOP07 in their simulations. This is mainly due to 
the radiative losses that are very efficient at $T \sim 10^5$~K and that cause a bump of $EM(T)$ at $T \sim 10^4$~K. 
We have verified this point by comparing simulations either with or without the radiative losses.}

\section{Discussion and conclusions} 
\label{discussion}

{In this work, we studied the scenario of a stellar jet impacting a molecular cloud from 
a HD modeling approach. In particular, we simulated supersonic stellar jets moving through 
ISM medium that is less dense than the jet encountering a dense molecular cloud. Hence, 
the jet becomes underdense inside the molecular cloud. From previous results (e.g. BOP07), 
X-ray emission is expected when the jet is lighter than the surrounding medium, while 
it is not when the jet is heavier than the ambient.}

{Our simulations explore the parameter
space: jet's velocity, ISM density and molecular cloud density. 
The scenario of a light jet
that impacts a dense cloud may be typical of some stellar jets observed in star-forming regions. 
In particular, this scenario is possibly occurring for HH~248. Our main goal is to investigate 
whether such an impact may produce X-ray emission from the jet.}

{Our results from the simulations indicate that jets with $v_\mathrm{jet} \gtrsim 
700$~km\,s$^{-1}$ can reach temperatures of several MK after impacting the dense 
molecular cloud. This result is in agreement with detection of X-ray emission from the 
jet. In contrast, low velocity jets do not reach such temperatures. We find a clear trend of
increasing temperature after impact with increasing the jet's velocity, independently 
of the density ratio between the jet and the dense molecular cloud.}

{We used these results to investigate the possible origin and evolution of HH~248, 
including the possibility of X-ray emission after encountering the high density region 
at the base of the Horsehead Nebula.} From our multiwavelength analysis 
of the {HH} object, we reveal the probable {driving source} of the jet, which is a highly absorbed 
{point-like source} detected only in a deep H-band image with the HST/WFC3. 
From the WFC3 image, we identify five dense knots that are not fully {aligned with WFC3~J05412764-0223143}. 
We used them to determine the parameters of the precession of the jet {associated with} HH~248. The 
fitted precession curve pass through the extended X-ray source named by \citet{lop13} as
Src.~12. We conclude that this X-ray source is the front shock of the jet that created HH~248. 

From our X-ray data analysis, we discard {the possibility that the extended X-ray 
source were part of the PSF of the classical T Tauri star V615~Ori, also detected 
in the {XMM-Newton} as $\sim 30\arcsec$}.
In particular, the column density determined for 
this source is one order of magnitude higher than that determined for V615~Ori by
\citet{lop13}. We propose a scenario for HH~248 in which the jet, which moves
through the ISM, reaches a region of much higher density. This region may be identified 
with the ionization front IC~434 that runs south from the NGC~2024 star-forming region 
and contains the Horsehead Nebula (Barnard~33). In this scenario, the jet, which is heavier 
than the ISM but lighter than the molecular cloud, emits in X-rays when it reaches the 
denser cloud. 
Hence, the X-ray emission from the jet {would be} produced far from the jet base, 
{contrarily to other cases (see Section~\ref{intro})}. In the classification 
given by \citet{bon10b}, luminous X-ray jets detected at large distances from their 
driving source seem to be associated with high-mass stars. In our case, the location of 
the X-ray emission from the jet {would not be} related to the mass of the driving 
source but to the change in the conditions of the ISM. 
This is an excellent test case to check the idea that detectable X-ray emission from 
stellar jets is produced if the jet travels through a denser environment (BOP07).
{From our results, we conclude that X-ray emission from the impact of a stellar 
jet with a dense cloud is feasible. Deep X-ray observations of HH objects in nearby star 
forming-regions may reveal other cases similar to HH~248.} New X-ray observations 
are needed to determine precisely some parameters of the X-ray counterpart of the jet.

\acknowledgments

The software used in this work was in part developed by the DOE-supported ASC/Alliance 
Center for Astrophysical Thermonuclear Flashes at the University of Chicago.
{HST data presented in this paper were obtained from the Mikulski Archive for Space 
Telescopes (MAST). STScI is operated by the Association of Universities for Research in 
Astronomy, Inc., under NASA contract NAS5-26555. Support for MAST for non-HST data 
is provided by the NASA Office of Space Science via grant NNX13AC07G and by other 
grants and contracts.} This work was supported by the Spanish Government under research projects 
AYA2011-29754-C03-01 and AYA2011-29754-C03-03. M. O. and J.F.A-C. 
acknowledge support by the Consejo Nacional de Investigaciones Cient\'ificas y T\'ecnicas 
(CONICET, Argentina). {Finally, we thank the referee for useful comments and 
suggestions.}




\clearpage

\end{document}